\documentclass[12pt]{iopart}
\usepackage{amssymb}
\usepackage[amssymb]{SIunits}
\usepackage{graphicx}
\usepackage{xspace}
\usepackage{bm}
\usepackage[colorlinks,urlcolor=blue]{hyperref}
\usepackage[all]{hypcap}
\newcommand*{\doi}[1]{\href{http://dx.doi.org/\detokenize{#1}}{\detokenize{#1}}}

\begin{document}

\title[RMPs for ELM suppression in ITER and their impact on divertor performance]{Heuristic predictions of RMP configurations for ELM suppression in ITER burning plasmas and their impact on divertor performance}

\author{H Frerichs${}^1$, J van Blarcum${}^1$, Y Feng${}^2$, L Li${}^3$, Y Q Liu${}^4$, A Loarte${}^5$, J-K Park${}^6$, R A Pitts${}^5$, O Schmitz${}^1$, S M Yang${}^7$}

\address{${}^1$ Department of Engineering Physics, University of Wisconsin - Madison, WI, USA}
\address{${}^2$ Max-Planck-Institut f\"ur Plasmaphysik, Greifswald, Germany}
\address{${}^3$ College of Science, Donghua University, Shanghai, China}
\address{${}^4$ General Atomics, San Diego, CA, USA}
\address{${}^5$ ITER Organization, St. Paul Lez Durance Cedex, France}
\address{${}^6$ Seoul National University, Korea}
\address{${}^7$ Princeton Plasma Physic Laboratory, Princeton, NJ, USA}

\ead{hfrerichs@wisc.edu}

\begin{abstract}
A subspace of resonant magnetic perturbation (RMP) configurations for edge localized mode (ELM) suppression is predicted for H-mode burning plasmas at 15 MA current and 5.3 T magnetic field in ITER.
Perturbation of the core plasma can be reduced by a factor of 2 for equivalent edge stability proxies, while the perturbed plasma boundary geometry remains mostly resilient.
The striation width of perturbed field lines connecting from the main plasma (normalized poloidal flux $< 1$) to the divertor targets is found to be significantly larger than the expected heat load width in the absence of RMPs.
This facilitates heat load spreading with peak values at an acceptable level below $10 \, \mega\watt \, \meter^{-2}$ on the outer target already at moderate gas fueling and low Ne seeding for additional radiative dissipation of the $100 \, \mega\watt$ of power into the scrape-off layer (SOL).
On the inner target, however, re-attachment is predicted away from the equilibrium strike point due to increased upstream heat flux, higher downstream temperature and less efficient impurity radiation.
\end{abstract}

\vspace{2pc}

\submitto{\NF}

\def\CmaxX{\ensuremath{C_{\textnormal{maxX}}}\xspace}
\def\CminM{\ensuremath{C_{\textnormal{minM}}}\xspace}
\def\CisoM{\ensuremath{C_{\textnormal{isoM}}}\xspace}
\def\Phirow#1{\ensuremath{\Phi_{\mathrm{#1}}}\xspace}
\def\xiX{\ensuremath{\boldsymbol \xi_X}\xspace}
\def\xiXrow#1{\ensuremath{\boldsymbol \xi_{X \mathrm{#1}}}\xspace}
\def\xiXnorm{\ensuremath{\xi_X^\ast}\xspace}
\def\xiM{\ensuremath{\boldsymbol \xi_M}\xspace}
\def\xiMnorm{\ensuremath{\xi_M^\ast}\xspace}
\def\kAt{\ensuremath{\kilo\ampere \textnormal{t}}\xspace}
\def\IRMP{\ensuremath{I_{\textnormal{RMP}}}\xspace}
\def\IRMPX{\ensuremath{I^X_{\textnormal{RMP}}}\xspace}
\def\Ith{\ensuremath{I_{\textnormal{th}}}\xspace}
\def\smax{\ensuremath{s_{\textnormal{max}}}\xspace}
\def\psiN{\ensuremath{\psi_N}\xspace}
\def\csepx{\ensuremath{c_\textnormal{sepx}}\xspace}
\def\qmax{\ensuremath{q_\textnormal{max}}\xspace}
\def\vec#1{\ensuremath{{\bf #1}}\xspace}

\section{Introduction}

Good particle and energy confinement is required for ITER to achieve and sustain burning plasma operation with a fusion gain of $Q = 10$ (i.e. where the internally generated fusion power exceeds the externally applied heating power by an order of magnitude), and the discovery of high-confinement (H-mode) operation \cite{Wagner1982, Wagner2007} has been essential for fusion applications and ITER in particular.
However, the reference regime exhibits so called type-I edge localized modes (ELMs) \cite{Zohm1996} (an MHD instability with fast loss of particles and energy from the plasma edge), and extrapolations to ITER have indicated that the ELM peak energy fluence to divertor targets can exceed acceptable values for lifetime considerations (surface melting) \cite{Federici2003a, Loarte2003, Gunn2017, Eich2017}.
A number of different ELM mitigating techniques are being investigated (see e.g. overview paper \cite{Evans2013a}).

\begin{figure}
\centering
\includegraphics[width=120mm]{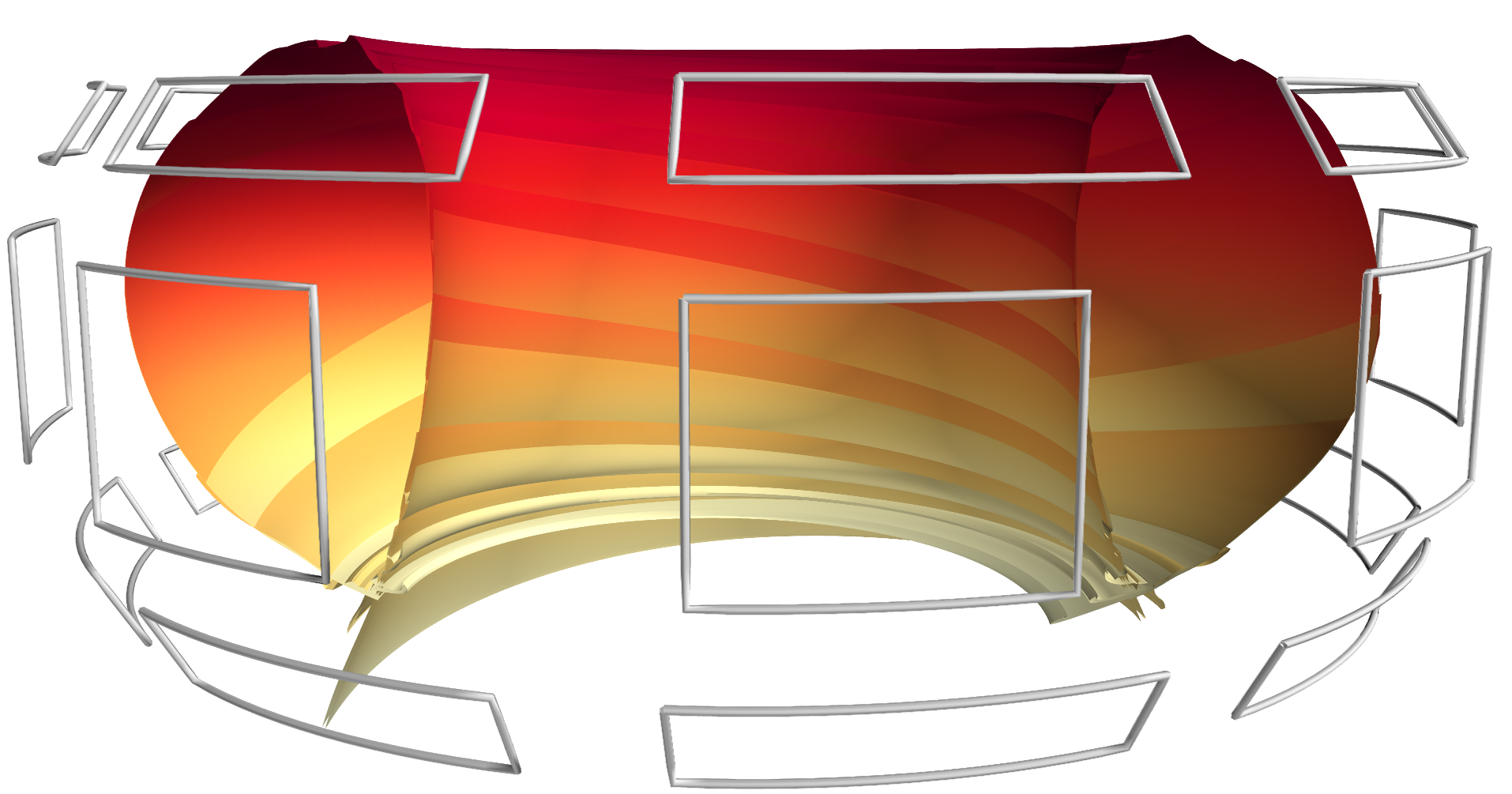}
\caption{Perturbed magnetic separatrix from application of RMPs (configuration \CmaxX from figure \ref{fig:scan2d_fixed_IRMP} for $\IRMP \, = \, 60 \, \kAt$). The geometry of the in-vessel ELM control coils is highlighted.}
\label{fig:separatrix3d}
\end{figure}

Suppression of ELMs by application of resonant magnetic perturbations (RMPs) has been pioneered at DIII-D \cite{Evans2004}, and it has since been successfully reproduced at KSTAR \cite{Jeon2012}, EAST \cite{Sun2016} and ASDEX Upgrade \cite{Suttrop2018}.
It is presently the main strategy for suppression of ELMs in ITER \cite{Lang2013, Loarte2014}.
Application of RMPs leads to helical corrugations of the magnetic separatrix as shown in figure \ref{fig:separatrix3d}, and those appear as lobes in camera images of the X-point region \cite{Kirk2012} and as non-axisymmetric striations of particle and heat fluxes onto divertor targets \cite{Evans2005a, Schmitz2008a, Ahn2010}.
The ITER divertor design \cite{Kukushkin2011, Pitts2019}, however, is based on 2-D (axisymmetric) SOLPS-4.3 and SOLPS-ITER plasma boundary modeling which neglects these effects.
This paper addresses implications of RMPs by 3-D modeling with EMC3-EIRENE \cite{Feng2004, Frerichs2021}.
Previous modeling of the pre-fusion power operation phase at lower power has shown that detachment onset occurs at lower upstream density, but at the expense of diverting heat flux to far SOL strike locations which are more difficult to detach \cite{Frerichs2020, Frerichs2021a}.
The remaining challenge is to verify the compatibility of RMPs for ELM suppression with a partially detached divertor plasma for safe handling of steady state power loads under burning plasma conditions with seeded impurities.

The stabilizing impact of RMPs on the plasma edge is associated with field penetration at the pedestal top \cite{Nazikian2015, Sun2016}, and this requires a certain level of perturbation depending on the edge safety factor $q_{95}$, density and plasma shape \cite{Hu2020, Hu2021}.
In the following we focus on the standard operating scenario for an ITER burning plasma with $15 \, \mega\ampere$ plasma current, $5.3 \, \tesla$ magnetic field, $q_{95} = 3.14$ and fusion gain of $Q = 10$.
Furthermore, we focus on perturbations with toroidal mode number $n = 3$ and evaluate the dependence on configuration parameters (for a comparison of divertor performance between $n = 3$ and $n = 4$ see \cite{VanBlarcum2024}).
As can be seen in figure \ref{fig:separatrix3d}, the in-vessel ELM control coils in ITER are arranged in three rows, and this offers the opportunity to align the perturbation from adjacent rows with the helical pitch of the magnetic field lines at the plasma edge.
This leaves a 5-dimensional configuration space of RMP parameters: the amplitudes of the current waveform in each of the three rows, as well as the relative phases between adjacent rows (barring other symmetry breaking effects, the global phase of the perturbation can be neglected).


Modeling of the impact of RMPs on ELMs requires time dependent, non-linear MHD models such as JOREK \cite{Huijsmans2015}, but those need a significant amount of computational resources and are unsuitable to evaluate a large number of configurations.
Therefore, for the purpose of scanning through the RMP parameter space, we rely on linear MHD models and figures of merit that correlate with ELM suppression, while referring to full MHD results only for a selected operating point as reference.
The aim of this paper is to evaluate the relation between these figures of merit for ELM suppression and divertor performance.
Specifically, we take the radial displacement of the plasma boundary surface near the X-point (referred to as X-point displacement) from linear, resistive MHD plasma response (MARS-F) as figure of merit for edge stability \cite{Liu2016, Liu2017, Li2017, Li2019}.
In section \ref{sec:scan2d}, we will map out RMP configurations for ELM suppression in ITER, and evaluate the relation to the magnetic footprint on the divertor targets as a proxy for particle and heat loads.
Then, in section \ref{sec:benchmark}, we revisit results for KSTAR where the ideal perturbed equilibrium code (IPEC) \cite{Park2007a} has been used to reproduce the experimentally observed operating window of RMP configurations with ELM suppression \cite{Park2018}.
We introduce code independent figures of merit, and demonstrate the robustness and compatibility of the two approaches.
Finally, we take particular RMP configurations of interest and follow up with 3-D plasma boundary simulations in section \ref{sec:heat_loads} in order to explore compatibility with divertor operating limits.


\section{RMP configurations for ELM suppression} \label{sec:scan2d}

With five free parameters, the RMP configuration space in ITER is vast.
The advantage of linear MHD models is that they are fast enough to map out the configuration space in order to identify operating points that appear beneficial for ELM suppression.
In this paper, we follow up on earlier results \cite{Li2019, Li2020} for the $Q = 10$ ITER scenario with $15 \, \mega\ampere$ current and $5.3 \, \tesla$ toroidal field, and further analyze the plasma response obtained from the MARS-F code \cite{Liu2000, Liu2010} which is based on a single fluid, linearized resistive MHD model.
Specifically, we pick one case based on a flow profile predicted by ASTRA transport simulations for $33 \, \mega\watt$ of NBI power, $20 \, \mega\watt$ ECRH power, Prandtl number $\Pr \, = \, 0.3$ and a ratio of the toroidal momentum to thermal confinement time of $\tau_\Phi / \tau_E \, = \, 1$.

\begin{figure*}
\centering
\includegraphics[width=0.9\textwidth]{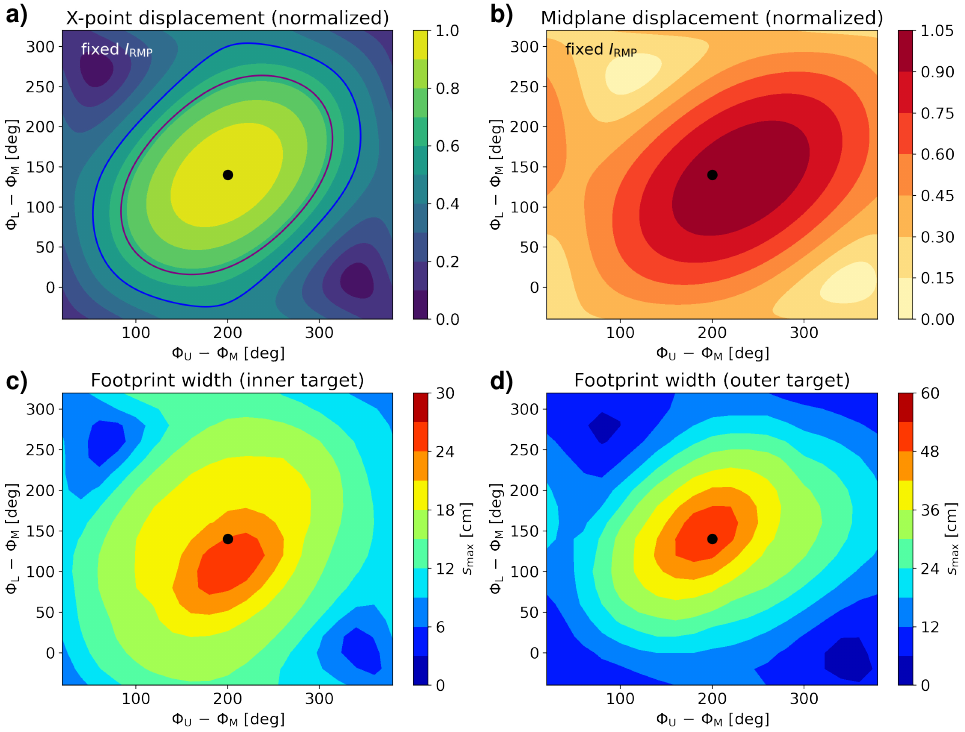}
\caption{Phase scan of the RMP waveform in the upper and lower row relative to the middle row at fixed amplitude \IRMP in all rows: (a) Normalized X-point displacement \xiXnorm, (b) normalized midplane displacement \xiMnorm, (c-d) magnetic footprint width \smax on the inner and outer target for $\IRMP = 60 \, \kAt$. The black dot highlights the configuration \CmaxX with $\max|\xiX|$. Contour lines are shown in (a) for $\xiXnorm = 1/2$ (blue) and 2/3 (purple) consistent with the threshold range for ELM suppression found in JOREK simulations \cite{Becoulet2022}.}
\label{fig:scan2d_fixed_IRMP}
\end{figure*}

\subsection{Configuration scan with fixed RMP current}

Despite initial success of the vacuum island overlap width (VIOW) as a proxy for ELM suppression \cite{Fenstermacher2008}, it is now accepted that plasma response plays a key role \cite{Lanctot2013, PazSoldan2015, Logan2016}.
A new figure of merit has been emerging: the plasma boundary displacement $|\xiX|$ near the X-point.
This ``X-point displacement`` is caused by the edge-peeling response \cite{Ryan2015, Liu2016}, and maximizing it is found to be best for ELM suppression.
Note that \xiX from MARS-F is a single complex number associated with toroidal mode $n$.
The plasma response is computed once for each row at $1 \, \kAt$, and the resulting \xiX from any current ($I_U, I_M, I_L$) and phase ($\Phirow{U}, \Phirow{M}, \Phirow{L}$) combination can then be obtained from a linear superposition

\begin{equation}
\xiX \, = \, I_U \, \xiXrow{U}^{(1 \, \kAt)} \, e^{i \, \Phirow{U}} \, + \, I_M \, \xiXrow{M}^{(1 \, \kAt)} \, e^{i \, \Phirow{M}} \, + \, I_L \, \xiXrow{L}^{(1 \, \kAt)} \, e^{i \, \Phirow{L}} \label{eq:xiX}
\end{equation}

of the upper (U), middle (M) and lower (L) rows.
As linear models cannot explain or predict a threshold value for ELM suppression, we can instead use the normalized magnitude

\begin{equation}
\xiXnorm \, = \, \frac{|\xiX|}{\max|\xiX|} \label{eq:xiXnorm}
\end{equation}

for picking the optimal operating point.
The magnitude only depends on the relative phases $\Phirow{U} \, - \, \Phirow{M}$ and $\Phirow{L} \, - \, \Phirow{M}$, and figure \ref{fig:scan2d_fixed_IRMP} (a) shows this dependence at fixed amplitude \IRMP in all rows.
Maximizing \xiXnorm is achieved by aligning the contributions to \xiX from each row, i.e.

\begin{equation}
\Phirow{U} \, - \, \Phirow{M} \Big|_{\textnormal{opt}} \, = \, \arg\left(\xiXrow{M}^{(1 \, \kAt)}\right) \, - \, \arg\left(\xiXrow{U}^{(1 \, \kAt)}\right)
\end{equation}

and the equivalent for the lower row.
An optimal RMP configuration is found at $\CmaxX \approx (200, 140) \, \deg$ as shown by the black dot in figure \ref{fig:scan2d_fixed_IRMP} (a).
It has been found that this operating point is insensitive to the assumed plasma toroidal flow \cite{Li2019, Li2020}.
It will be the baseline for the following analysis.

Recently, identification of this optimal RMP configuration has motivated non-linear MHD simulations with the JOREK core in order to verify ELM suppression in ITER \cite{Becoulet2022}.
These simulations have predicted that the threshold for ELM suppression is in the range of $\IRMP \, = \, 45 -- 60 \, \kAt$.
This is somewhat larger than the value of $25 -- 30 \, \kAt$ predicted by TM1 based on a $15 \, \%$ pedestal pressure reduction \cite{Hu2021a} (TM1 is another non-linear time-dependent two-fluid MHD code with cylindrical geometry, albeit with circular cross-section, and coupled to GPEC \cite{Park2017} for boundary conditions).
The TM1 results are, however, very sensitive to $q_{95}$ (see figure 9 in \cite{Hu2021a}).
In the following, we will leverage the threshold range identified by JOREK and refer to the lower bound ($45 \, \kAt$) as {\it optimistic} scenario and to the upper bound ($60 \, \kAt$) as {\it conservative} scenario.
With $\max|\xiX|$ evaluated for the hardware limit of $90 \, \kAt$ in ITER, the threshold \xiXnorm for ELM suppression is 1/2 (optimistic) or 2/3 (conservative).
The blue and purple contours in figure \ref{fig:scan2d_fixed_IRMP} (a) show the corresponding boundary of the configuration space that supports ELM suppression.

Qualitatively, ELM suppression by RMPs requires {\it enough} perturbation at the plasma edge, but at the same time avoiding {\it too much} perturbation in the core in order to avoid mode locking.
The latter imposes an upper threshold for the applied RMP strength, and this is found to be a particularly strict condition for low $n$ RMPs in KSTAR \cite{In2017, Park2018, In2019a}.
A combined approach for maximizing the edge impact and minimizing the core impact allows for optimal ELM control \cite{Yang2020, Kim2023}.
It has been found that the core-kink response is linked to a large displacement $|\xiM|$ of the plasma boundary at the outboard midplane \cite{Liu2011, Liu2016, Liu2017}, and in the following we will use this as a figure of merit for core stability.
For the optimization approach below, we will normalize \xiM to its value at \CmaxX and $\IRMP = 90 \, \kAt$:

\begin{equation}
\xiMnorm \, = \, \frac{|\xiM|}{|\xiM|_{\max|\xiX|}} \label{eq:xiMnorm}
\end{equation}

where \xiM is obtained from a linear superposition similar to (\ref{eq:xiX}).
Note that values of \xiMnorm can exceed 1 in this definition, unlike \xiXnorm.
The phase dependence of \xiMnorm at fixed \IRMP is shown in figure \ref{fig:scan2d_fixed_IRMP} (b).
It can be seen in comparison to figure \ref{fig:scan2d_fixed_IRMP} (a) that large \xiXnorm tend to imply large \xiMnorm.
This shows that some trade-off between \xiXnorm and \xiMnorm may need to be considered (as already suggested in \cite{Li2019}).
At KSTAR, an empirical limit for \xiMnorm is obtained from the standard $90 \, \deg$ configuration for $n = 1$ RMPs, and this implies that large parts of the RMP configuration space are excluded \cite{In2017, Park2018}.
For $n = 3$ RMPs in ITER, this may not be as restrictive.
Without a known limit value for ITER, we can at least map out which regions of the RMP configuration space are less susceptible for core-kink response than others.
We will follow up on this below in section \ref{sec:scan2d_fixed_xiX}.

Another important effect to consider is the impact of RMPs on particle and heat loads on divertor targets.
Application of RMPs lead to helical corrugations of the magnetic separatrix, and those guide field lines from the interior to the divertor targets.
Transport along those perturbed field lines results in non-axisymmetric striations of particle and heat loads on divertor targets which follow the geometry of the perturbed separatrix.
Therefore, the magnetic footprint can be used as a proxy for characterization of particle and heat load striations.
The FLARE code \cite{Poster-APS2015-Frerichs} is used to trace field lines from the divertor targets and to track their deepest connection into the plasma $\min\psiN|_{\textnormal{fieldline}}$ along the way (\psiN is the normalized poloidal magnetic flux).
The footprint width \smax is then defined as the maximum distance from the equilibrium strike point with $\min\psiN|_{\textnormal{fieldline}} < 1$.
Despite the linear superposition of the MARS-F plasma response from each row, field line tracing and the computation of \smax is of non-linear nature.
Figure \ref{fig:scan2d_fixed_IRMP} (c) and (d) shows the resulting \smax on the inner and outer target, respectively, for $\IRMP \, = \, 60 \, \kAt$.
Each magnetic footprint is sampled with a resolution of $1 \, \deg$ toroidally and $1 \, \centi\meter$ along the target, and the 2-D phase scan is conducted with a resolution of $20 \, \deg$.
It can be seen that large \xiXnorm tends to imply large \smax both on the inner and outer target.
This is consistent with computational observations at DIII-D \cite{Munaretto2022} based on the high field side plasma response as proxy for ELM suppression.
Furthermore, it can be seen that the magnetic footprints on the outer target are about a factor of 2 larger than on the inner target.
This is a consequence of the target geometry in relation to the shape of the equilibrium: flux surfaces have a shallower (poloidal) incident angle on the outer target which implies a larger distance from the equilibrium strike point for the same \psiN.

For attached plasmas, RMPs can spread the heat loads over a wider area, depending on the magnetic footprint width \smax in relation to the reference heat load width without RMPs.
With an upstream heat flux width of $\lambda_q = 3.4 \, \milli\meter$ \cite{Pitts2019} (or below \cite{Eich2013}), and a flux expansion of $f_x \approx \, 6$, it can be expected that the perturbed configuration will significantly impact the divertor plasma in ITER.
The ITER divertor, however, is required to operate in a partially detached state in order to remain below an acceptable level of about $10 \, \mega\watt \, \meter^{-2}$.
Previous EMC3-EIRENE simulations for the pre-fusion power operation phase at lower heating power have shown that while detachment onset occurs at lower upstream density, heat flux is shifted along the helical corrugations of the separatrix into the far SOL strike locations which remain attached towards higher upstream density \cite{Frerichs2020, Frerichs2021a}.
Thus, maximizing \smax may not be the best approach for ITER, if power dissipation from impurity seeding turns out to be insufficient for reducing heat loads at far SOL strike points.
We will follow up on this in section \ref{sec:heat_loads} for the configurations of interest identified below.

\subsection{Configuration scan with fixed \xiXnorm} \label{sec:scan2d_fixed_xiX}

\begin{figure*}
\centering
\includegraphics[width=0.9\textwidth]{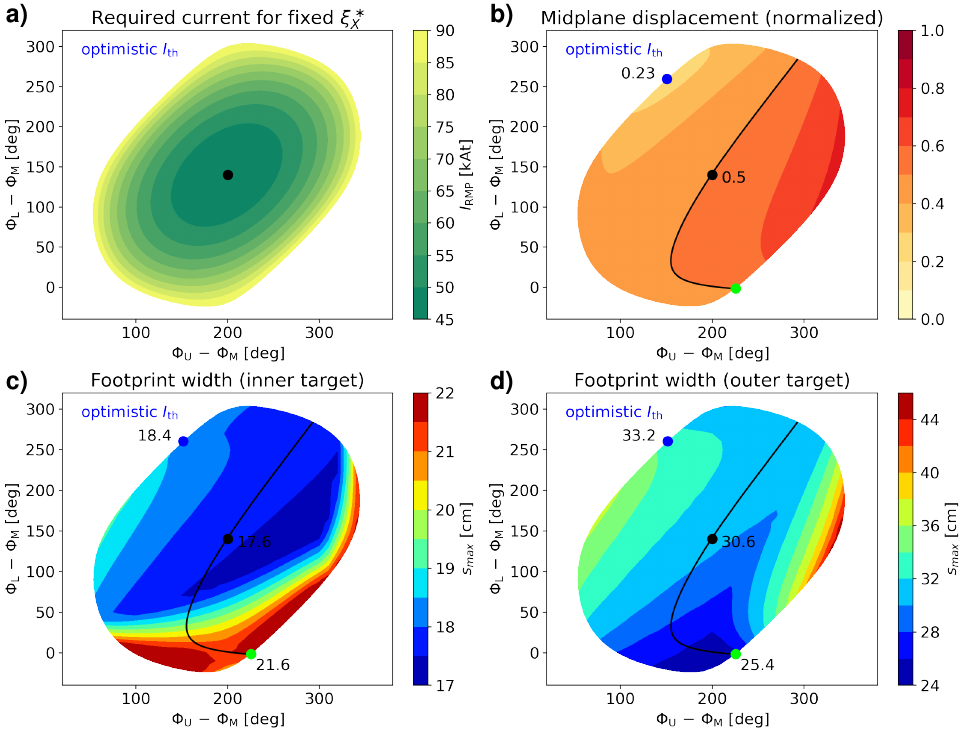}
\caption{Phase scan of the RMP waveform in the upper and lower row with relative to the middle row at fixed \xiXnorm for the optimistic threshold at $\Ith = 45 \, \kAt$: (a) Required current \IRMP, (b) normalized midplane displacement \xiMnorm, (c-d) magnetic footprint width \smax on the inner and outer target. The black dot marks the reference configuration \CmaxX from figure \ref{fig:scan2d_fixed_IRMP}, and the blue dot highlights the configuration $\CminM^{(45)}$ with minimal \xiMnorm that is compatible with the maximum available \IRMP. The black line marks the iso-\xiMnorm contour through \CmaxX, and the green dot highlights the configuration $\CisoM^{(45)}$ with minimal \smax on the outer target along that contour.}
\label{fig:scan2d_fixed_xiX_Ith45}
\end{figure*}

Based on the threshold(s) identified by JOREK simulations, we will now look at configurations with the same \xiXnorm in the anticipation that they imply equivalent ELM suppression.
Knowing the phase dependence of \xiXnorm as shown in figure \ref{fig:scan2d_fixed_IRMP} (a), we can define the RMP current

\begin{equation}
\IRMP(\Ith, \Delta\Phirow{UM}, \Delta\Phirow{LM}) \, = \, \frac{\Ith}{\xiXnorm(90 \, \kAt, \Delta\Phirow{UM}, \Delta\Phirow{LM})}
\end{equation}

that is required for each phase combination in order to match the \xiXnorm of the ELM suppression threshold \Ith at \CmaxX.
Figure \ref{fig:scan2d_fixed_xiX_Ith45} (a) shows the optimistic scenario $\IRMP (\Ith = 45 \, \kAt)$ which reproduces the value of \Ith at \CmaxX (black dot).
Configurations which exceed the $90 \, \kAt$ limit are excluded (white), but about half ($52.2 \, \%$) of the phase combinations support ELM suppression.

Figure \ref{fig:scan2d_fixed_xiX_Ith45} (b) shows the corresponding \xiMnorm as figure of merit for the core response.
As the normalization of \xiMnorm in (\ref{eq:xiMnorm}) is based on the \CmaxX configuration at full current, a value of 0.5 is found here for this configuration in the optimistic \Ith scenario.
It can be seen that the impact on the core plasma can be reduced by a factor of 2 for the configuration $\CminM^{(45)} = (151, 261) \, \deg$ (blue dot) with respect to the reference configuration \CmaxX.
The maximum of \xiMnorm is below 0.74, but at this point we do not have an upper limit for core stability (which in KSTAR significantly reduced the set of viable configurations).
Overall, $29.5 \, \%$ of the phase combinations support ELM suppression at $\xiMnorm \, \le \, 1/2$ (left of the black line).

The corresponding magnetic footprint widths on the inner and outer divertor targets are shown in figure \ref{fig:scan2d_fixed_xiX_Ith45} (c-d).
A much smaller variation of \smax is found now that all configurations have equal \xiXnorm.
Thus, decoupling of the footprint width from ELM control may be challenging, as already pointed out by Munaretto et al. \cite{Munaretto2022}.
However, the 2-D phase scan here shows that the three row setup of ITER (and KSTAR) supports fine tuning that would otherwise be severely limited with a two row setup such as DIII-D.
Here, a higher resolution of $0.1 \, \deg$ and $0.1 \, \centi\meter$ is used to sample field lines for the 3 configurations of particular interest (highlighted by the black, blue and green dots).
The magnetic footprints at \CmaxX are smaller here ($17.6 \, \centi\meter$ on the inner target and $30.6 \, \centi\meter$ on the outer target) than in figure \ref{fig:scan2d_fixed_IRMP} (c-d) because of the lower $\IRMP \, = \, 45 \, \kAt$ applied here.
A minor increase to $18.4 \, \centi\meter$ (inner target) and $33.2 \, \centi\meter$ (outer target) is found for $\CminM^{(45)}$ with improved core stability.
On the other hand, a somewhat larger variation of \smax is found along the black iso-\xiMnorm line where configurations are equivalent in both ELM suppression and core stability.
In particular, for the configuration $\CisoM^{(45)} = (225, -2) \, \deg$ (green dot), \smax is increased to $21.6 \, \centi\meter$ ($+ 23 \, \%$) on the inner target while \smax is decreased to $25.4 \, \centi\meter$ ($- 17 \, \%$) on the outer target compared to \CmaxX.

\begin{figure*}
\centering
\includegraphics[width=0.9\textwidth]{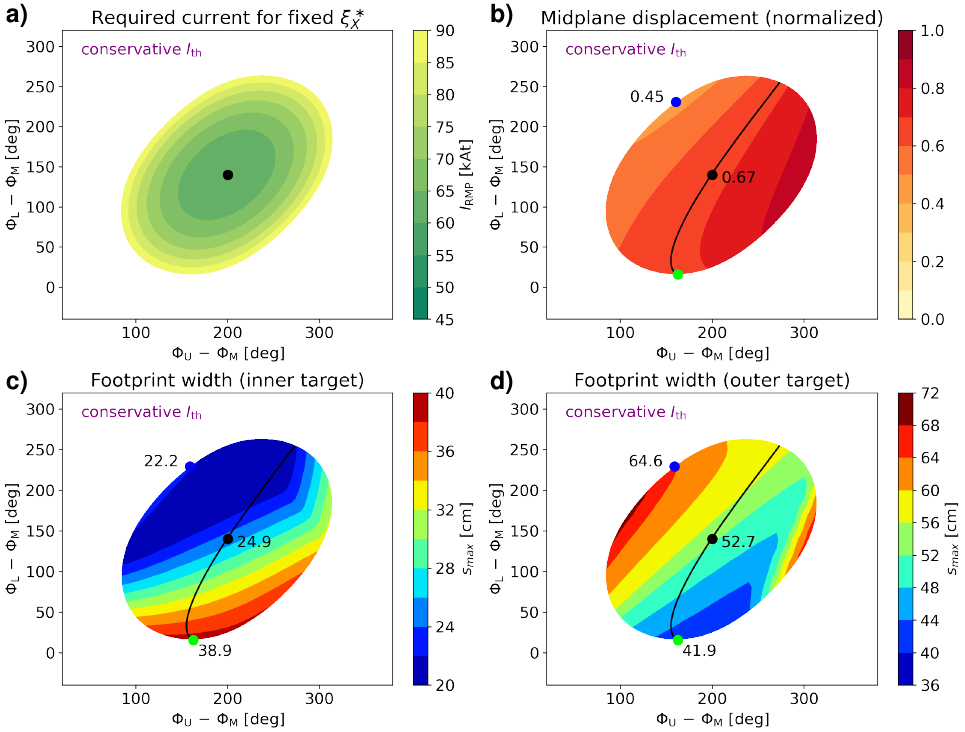}
\caption{Phase scan of the RMP waveform in the upper and lower row with relative to the middle row at fixed \xiXnorm for the optimistic threshold at $\Ith = 60 \, \kAt$: (a) Required current \IRMP, (b) normalized midplane displacement \xiMnorm, (c-d) magnetic footprint width \smax on the inner and outer target. The black dot marks the reference configuration \CmaxX from figure \ref{fig:scan2d_fixed_IRMP}, and the blue dot highlights the configuration $\CminM^{(60)}$ with minimal \xiMnorm that is compatible with the maximum available \IRMP. The black line marks the iso-\xiMnorm contour through \CmaxX, and the green dot highlights the configuration $\CisoM^{(60)}$ with minimal \smax on the outer target along that contour.}
\label{fig:scan2d_fixed_xiX_Ith60}
\end{figure*}

The conservative scenario for $\Ith = \, 60 \, kAt$ is shown in figure \ref{fig:scan2d_fixed_xiX_Ith60}.
A smaller set of configurations ($31.6 \, \%$) supports ELM suppression compared to the optimistic scenario due to the higher \Ith at \CmaxX here.
Nevertheless, figure \ref{fig:scan2d_fixed_xiX_Ith60} (b) shows that a significant reduction of core perturbation is still possible with an optimum of $\xiMnorm = 0.45$ at $\CminM^{(60)} = (159, 229) \, \deg$ ($-32 \, \%$ relative to \CmaxX).
Overall, $16.8 \, \%$ of the phase combinations support ELM suppression with improved core stability ($\xiMnorm < 2/3$).
Magnetic footprints in figure \ref{fig:scan2d_fixed_xiX_Ith60} (c-d) are larger than in figure \ref{fig:scan2d_fixed_xiX_Ith45} (c-d) consistent with the larger \xiXnorm of the conservative scenario.
For the reference point \CmaxX, $\smax \, = \, 24.9 \, \centi\meter$ on the inner target and $\smax \, = \, 52.7 \, \centi\meter$ on the outer target.
The same trend of shifting the footprint size from the outer to the inner target is found when moving along the black iso-\xiM line towards $\CisoM^{(60)} = (162, 16) \, \deg$ (green dot).
For the improved core configuration $\CminM^{(60)}$ (blue dot), the footprint increases to $\smax \, = \, 64.6 \, \centi\meter$ ($+23 \, \%$) on the outer target.
It should be noted that the extent of the beveled tungsten monoblock region (the dedicated high heat flux region) on the outer vertical target is only $57.9 \, \centi\meter$ (for this equilibrium).
This implies that too much optimization for core perturbation may not be compatible with tolerable divertor heat loads.


\section{Figures of merit} \label{sec:benchmark}

The optimization of RMP configurations for ELM suppression relies on robust figures of merit.
At KSTAR, the experimentally observed operating window of RMP configurations with ELM suppression has been reproduced by linear, ideal MHD plasma response \cite{Park2018}.
Specifically, figures of merit for the edge ($\mathcal{E}$) and for the core ($\mathcal{C}$) have been constructed from the perturbed flux $\Phi_{mn}$:

\begin{equation}
\mathcal{E}, \mathcal{C} \, = \, \sqrt{\frac{1}{N} \, \sum \, |\Phi_{mn}|^2} \label{eq:Nsum}
\end{equation}

where the sum is taken over $N$ rational surfaces that have to be properly chosen for each of the two figures.
For $n = 1$ RMPs in KSTAR, the $m = 5, 6$ resonances are located in the edge and the $m = 2, 3$ resonances are located in the core.
Viable configurations require $\mathcal{E} \, \ge \, \mathcal{E}_{\textnormal{th}}$ for ELM suppression and $\mathcal{C} \, < \, \mathcal{C}_{\textnormal{crit}}$ to avoid disruptive core field penetration.
The critical values $\mathcal{E}_{\textnormal{th}}$ and $\mathcal{C}_{\textnormal{crit}}$ are determined from the empirical ELM suppression window in the $90 \, \deg$ standard configuration from $\IRMP \, = \, 1.8 \, \kilo\ampere$ to $2.1 \, \kilo\ampere$.

\begin{figure*}
\centering
\includegraphics[width=1.0\textwidth]{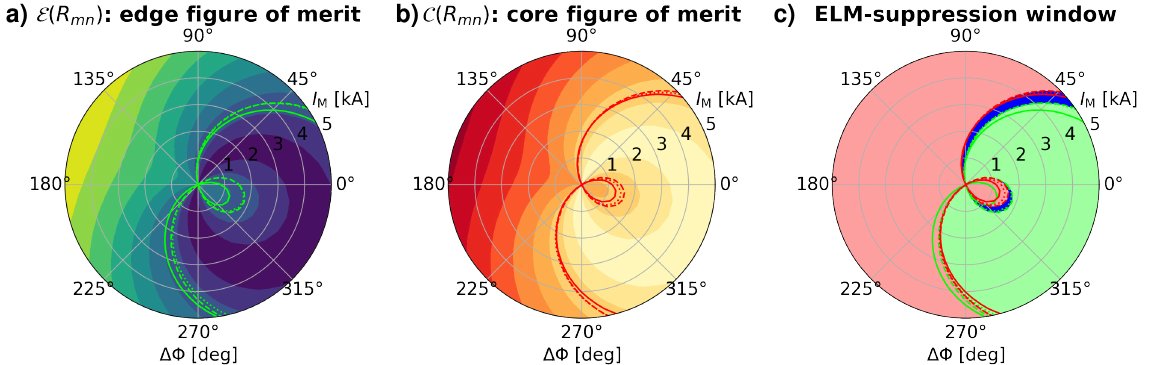}
\caption{(a) Edge figure of merit with threshold for ELM suppression (green), (b) core figure of merit with stability limit (red), and (c) operation window for ELM suppression (blue).
Contour lines are based on $R_{mn}$ (dashed) and $\varepsilon_{m+1 \, n}$ (dotted) in comparison to the $\Phi_{mn}$ reference (solid) from Park et. al \cite{Park2018}.
Here, $\Delta \Phi = \Phi_M \, - \, \Phi_L \, = \, \Phi_U \, - \, \Phi_M$ while $I_U \, = \, I_L \, = 5 / \sqrt{2} \, \kilo\ampere$ is kept fixed and $I_M$ is scanned.}
\label{fig:proxy_benchmark_KSTAR}
\end{figure*}

It should be noted that the $\Phi_{mn}$ from IPEC do not represent the resulting resonant fields (which are screened in ideal MHD), but rather their values when the singular currents at the resonances were to dissipate.
As such, $\Phi_{mn}$ is specific output of IPEC.
In order to link those results to ours - and to demonstrate the robustness of these figures of merit - we introduce a code independent approach below.
The Fourier transformed perturbed flux

\begin{equation}
\varepsilon_{mn}(\psi) \, = \, \frac{1}{\left( 2 \pi \right)^2} \, \oint \!d\theta d\varphi \, \mathcal{J} \, \frac{\delta\vec{B} \cdot \nabla \psi}{||\nabla \psi||} \, e^{-i \left(m \theta \, - \, n \varphi\right)} \label{eq:epsmn}
\end{equation}

is computed in a post-processing step from the perturbation field $\delta\vec{B}$ ($\mathcal{J}$ is the flux surface Jacobian in straight field line coordinates $\theta$ and $\varphi$ for the equilibrium poloidal flux $\psi$).
The FLARE code \cite {Poster-APS2015-Frerichs} has interfaces for $\delta\vec{B}$ from IPEC, MARS-F and other plasma response codes, and can evaluate (\ref{eq:epsmn}) in support of field line tracing.
The magnitude of the resonant field $|\varepsilon_{mn}(\psi_{mn})|$ for $q(\psi_{mn}) \, = \, m/n$ is typically small due to plasma response (or zero in ideal MHD).
For the new figures of merit, we replace $\Phi_{mn}$ in (\ref{eq:Nsum}) with either the kink amplitude $|\varepsilon_{m+1 \, n}(\psi_{mn})|$, or with the resonant response

\begin{equation}
R_{mn} \, = \, \frac{|\varepsilon_{mn}'|}{m} \, \frac{q}{q'} \Bigg|_{\psi_{mn}} \label{eq:Rmn}
\end{equation}

as a measure for the kinking of the resonant surface that can be attributed to the plasma response.
Derivatives ($'$) in (\ref{eq:Rmn}) are taken with respect to the radial direction $\psi$.
Figure \ref{fig:proxy_benchmark_KSTAR} shows an RMP configuration scan for the new edge and core figures of merit based on $R_{mn}$.
Contour lines for critical values are also shown for figures of merit based on $\varepsilon_{m+1 \, n}$ (dotted) and for the $\Phi_{mn}$ reference (solid).
Note that the RMP configuration scan here is different from figures \ref{fig:scan2d_fixed_IRMP} - \ref{fig:scan2d_fixed_xiX_Ith60}, but it is chosen in order to match the one in \cite{Park2018} as baseline for this benchmark.
It can be seen that the simultaneous constraints on $\mathcal{E}$ (left of green lines) and $\mathcal{C}$ (right of red lines) result in a narrow window of suitable RMP configurations for ELM suppression (blue).
In particular, it can be seen that the new figures of merit based on $R_{mn}$ and $\varepsilon_{m+1 \, n}$ are consistent with the IPEC ones based on $\Phi_{mn}$.

\begin{figure}
\centering
\includegraphics[width=80mm]{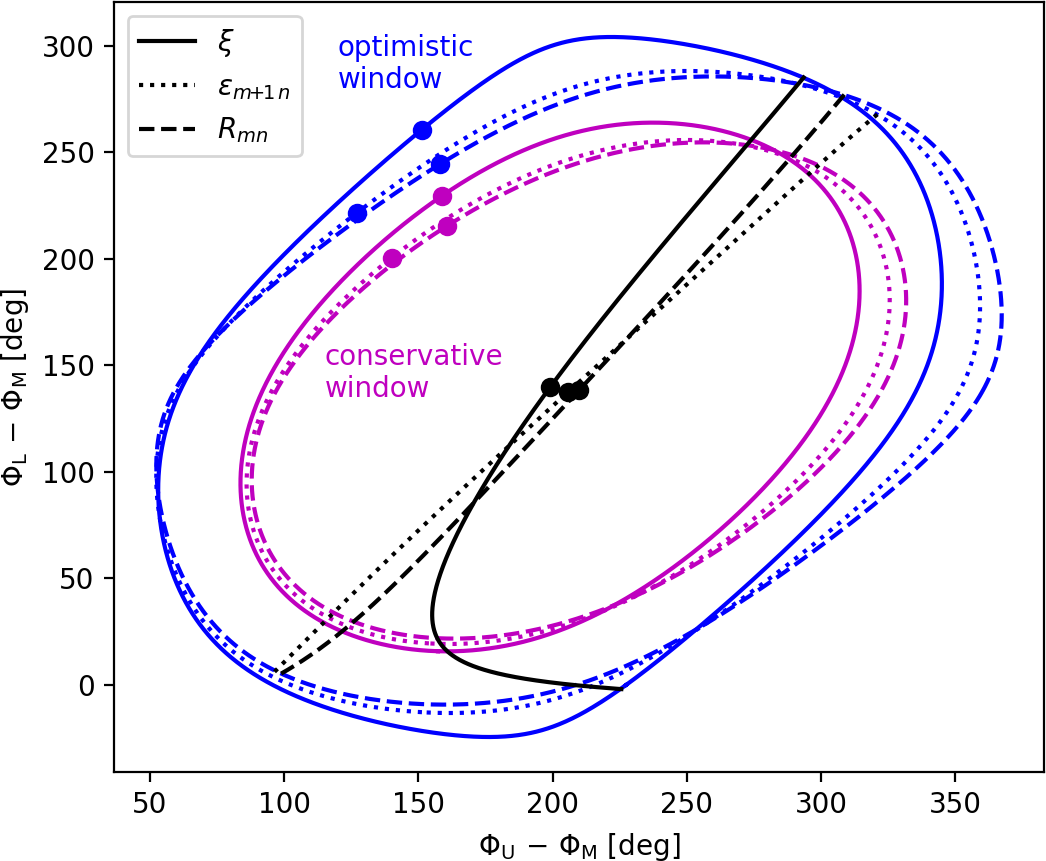}
\caption{Benchmark of figures of merit for ELM suppression. Blue and purple lines represent the boundary of viable configurations for an optimistic and conservative threshold, respectively. Black dots represent the configuration with maximum edge figure of merit at fixed RMP current, and black lines represent configurations with equivalent core perturbation at fixed edge figure of merit. Colored dots represent configurations with minimal core perturbation.}
\label{fig:proxy_benchmark_ITER}
\end{figure}

In order to improve confidence in the previously identified RMP configurations for ELM suppression in ITER, figure \ref{fig:proxy_benchmark_ITER} shows the results from section \ref{sec:scan2d} in comparison to the new figure of merit.
For $n = 3$ RMPs in ITER, $\mathcal{E}$ is determined from the edge resonances $m = 9, 10$ and 11, while $\mathcal{C}$ is determined from the core resonances $m = 4, 5$ and 6.
It can be seen that our earlier results based on the plasma displacement at the X-point and at the midplane are reproduced by the new figures of merit based on the kink amplitude $\varepsilon_{m+1 \, n}$ and the resonant response $R_{mn}$.
In particular, the black dots in figure \ref{fig:proxy_benchmark_ITER} mark the {\it optimal} configuration obtained from maximizing each of the edge figures of merit at fixed RMP current \IRMP.
By rescaling \IRMP to match the edge perturbation of the {\it optimal} configurations, the blue and purple lines highlight the boundary of viable configurations for ELM suppression for an optimistic threshold of $45 \, \kAt$ and a conservative threshold of $60 \, \kAt$, respectively.
At this fixed level of edge perturbation, the blue and purple dots mark the configurations with minimal core perturbation.
Furthermore, it is shown that similar contour lines (black) are found for configurations with both equivalent core and edge perturbation.
Even though some deviation may be noted for the optimistic window, this has little impact on the corresponding magnetic footprints in figure \ref{fig:scan2d_fixed_xiX_Ith45} (c,d).


\section{Impact on divertor heat loads} \label{sec:heat_loads}

\begin{figure*}
\centering
\includegraphics[width=0.5\textwidth]{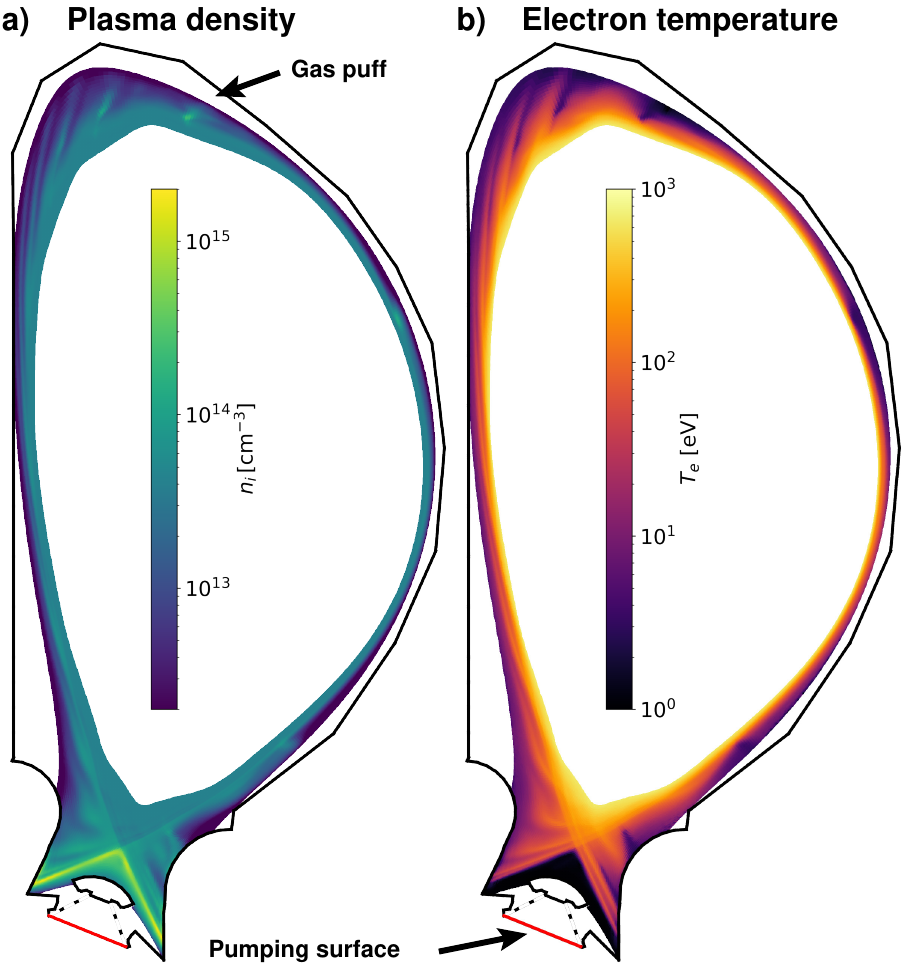}
\caption{Cross section of the plasma density (a) and electron temperature (b) at $\varphi \, = \, 0 \, \deg$ for the \CmaxX RMP configuration at $\IRMP \, = \, 45 \, \kAt$.}
\label{fig:rzslice_plasma}
\end{figure*}

Field line tracing is a relatively cheap method to determine where particle and heat loads would be expected from RMP application, but ultimately, 3D plasma edge and boundary modeling is required.
The EMC3-EIRENE code \cite{Feng2017} combines EMC3 \cite{Feng1999} for the fluid edge plasma with EIRENE \cite{Reiter2005} for a kinetic neutral particle transport.
Reactions include ionization, charge exchange, recombination, molecular processes such as dissociation, ion conversion and elastic $D_2 \, - \, D^+$ collisions, and neutral-neutral collisions (in BGK approximation).
The edge plasma model is a time-independent, simplified version of Braginskii's fluid equations for particle, momentum and energy balance.
Quasi neutrality $n_e = n_i$ is assumed, along with no plasma current (which implies $u_e = u_i$).
Classical heat conductivity $\kappa_e \sim T_e^{5/2}$ and $\kappa_i \sim T_i^{5/2}$ is assumed for electrons and ions, respectively, but cross-field particle and energy transport is assumed to be anomalous with model parameters $D$, $\chi_e$ and $\chi_i$.
Impurity transport with $T_Z \, = \, T_i$ is included in an approximation for small concentrations $Z^2 \, n_Z \, \ll \, n_e$, but corresponding radiation losses are coupled to the energy balance for electrons.
The balance equations for $n_i, u_i, T_i, T_e$ and $n_Z$ are cast into a Fokker-Planck form which is suitable for a Monte Carlo method \cite{Feng2000}.
A reversible field line mapping algorithm is applied for fast reconstruction of field lines from the 3D mesh \cite{Feng2005}, which is block-structured for applications in poloidal divertor geometry \cite{Frerichs2010}.

Simulations have been conducted for each of the three configurations of interest (\CmaxX, \CminM and \CisoM) from figure \ref{fig:scan2d_fixed_xiX_Ith45} and \ref{fig:scan2d_fixed_xiX_Ith60} for both \Ith scenarios.
All simulations have been conducted with the same model parameters: $100 \, \mega\watt$ input power into the edge, $\textnormal{D}_2$ gas puffing of $5 \, \cdot 10^{22} \, \second^{-1}$ from the top, core fueling of $9.1 \, \cdot 10^{21} \, \second^{-1}$, $100 \, \%$ recycling of $D^+$ on material surfaces, $0.72 \, \%$ absorption (pumping) of D neutrals below the divertor dome, anomalous cross-field transport of $D \, = \, 0.3 \, \meter^2 \, \second^{-1}$ and $\chi_e \, = \, \chi_i \, = \, 1 \, \meter^2 \, \second^{-1}$, and a Ne concentration of $\csepx \, = \, 0.8 \, \%$ at the separatrix.
This is consistent with the range of parameters used in SOLPS-4.3 and SOLPS-ITER simulations for the physics basis of the ITER divertor \cite{Pitts2019}.
The position of the gas puff and the pumping location is marked in figure \ref{fig:rzslice_plasma}.
Self consistent impurity throughput is not implemented in EMC3-EIRENE at this point.
Rather, impurity recycling of $99 \, \%$ is applied on material surfaces, and the source strength from Ne gas puffing is adjusted in order to maintain a given concentration \csepx evaluated in the outermost cell layer of the core zone of the computational mesh.

\begin{figure*}
\centering
\includegraphics[width=\textwidth]{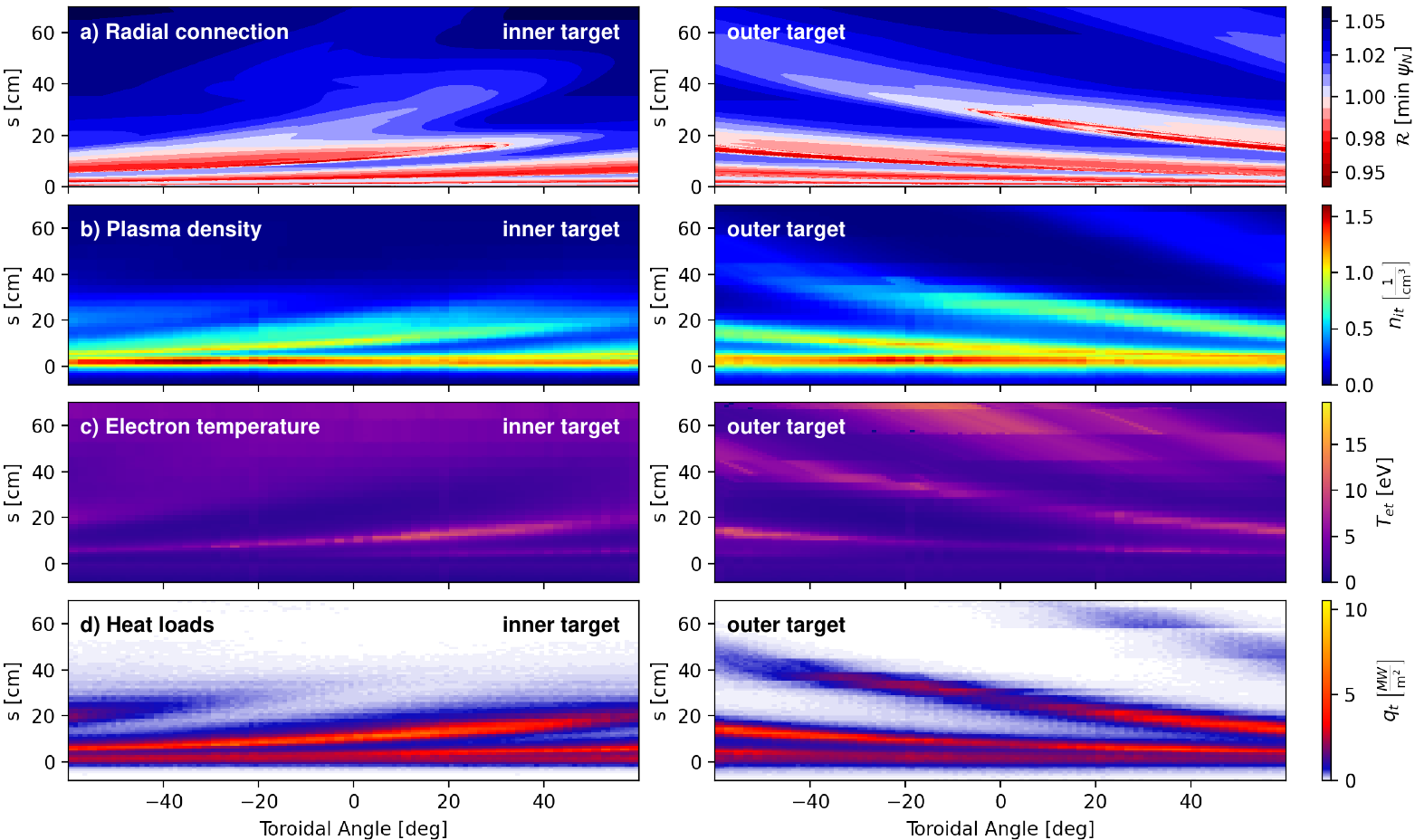}
\caption{Magnetic footprint and plasma conditions on the inner divertor target (left column) and outer divertor target (right column) for the \CmaxX configuration at $\IRMP \, = \, 45 \, \kAt$: (a) radial field line connection, (b) plasma density, (c) electron temperature, and (d) heat loads.
Color bars are chosen to accommodate the value range in figure \ref{fig:footprint_xiXmax60}.}
\label{fig:footprint_xiXmax45}
\end{figure*}

The plasma density and electron temperature for the \CmaxX RMP configuration at $\IRMP \, = \, 45 \kAt$ are shown in figure \ref{fig:rzslice_plasma}.
The lobe structure of the perturbed separatrix is clearly visible in the SOL plasma.
It can be seen that the density below the X-point is of the order $10^{15} \, \centi\meter^{-3}$.
This is about 25 times higher than the value of $4.2 \cdot 10^{13} \, \centi\meter^{-3}$ at the last closed flux surface.
The high density facilitates low temperatures in front of the divertor targets, but more so near the equilibrium strike point.
Plasma conditions at the divertor targets are shown in figure \ref{fig:footprint_xiXmax45} (b-d) in comparison to the magnetic footprint in (a) where red colors indicate field lines connecting from the plasma interior ($\psiN < 1$).
The density peak is located within a few $\centi\meter$ from the equilibrium strike point as a result of recycling in the private flux region (favored by the field line incident on the vertical targets), and the temperature drops below $2 \, \electronvolt$ there with heat loads of about $2 -- 3 \, \mega\watt \, \meter^{-2}$.
Along the lobe strike locations further out in the SOL ($s \gtrsim 10 \, \centi\meter$), on the other hand, the density is lower - but still in the range $0.5 -- 1 \, \cdot 10^{15} \, \centi\meter^{-3}$ - and the temperature remains above $10 \, \electronvolt$.
Heat loads of up to $6 \, \mega\watt \, \meter^{-2}$ can be found, which is still at an acceptable level.
This is significantly different from earlier results at lower upstream density \cite{Schmitz2016} where the plasma was still attached with peak loads of over $20 \, \mega\watt \, \meter^{-2}$ - but only because at that time EMC3-EIRENE was not able to handle partially detached conditions with strong neutral gas interactions.
Meanwhile, recombination has been implemented into EMC3-EIRENE, and the algorithm has been improved which now permits convergence at low divertor temperatures.
Comparisons to SOLPS-ITER show good agreement for the pre-fusion power operation phase \cite{Frerichs2019}, and we will follow up with a benchmark of impurity seeding at higher power at a later point.

\begin{figure*}
\centering
\includegraphics[width=\textwidth]{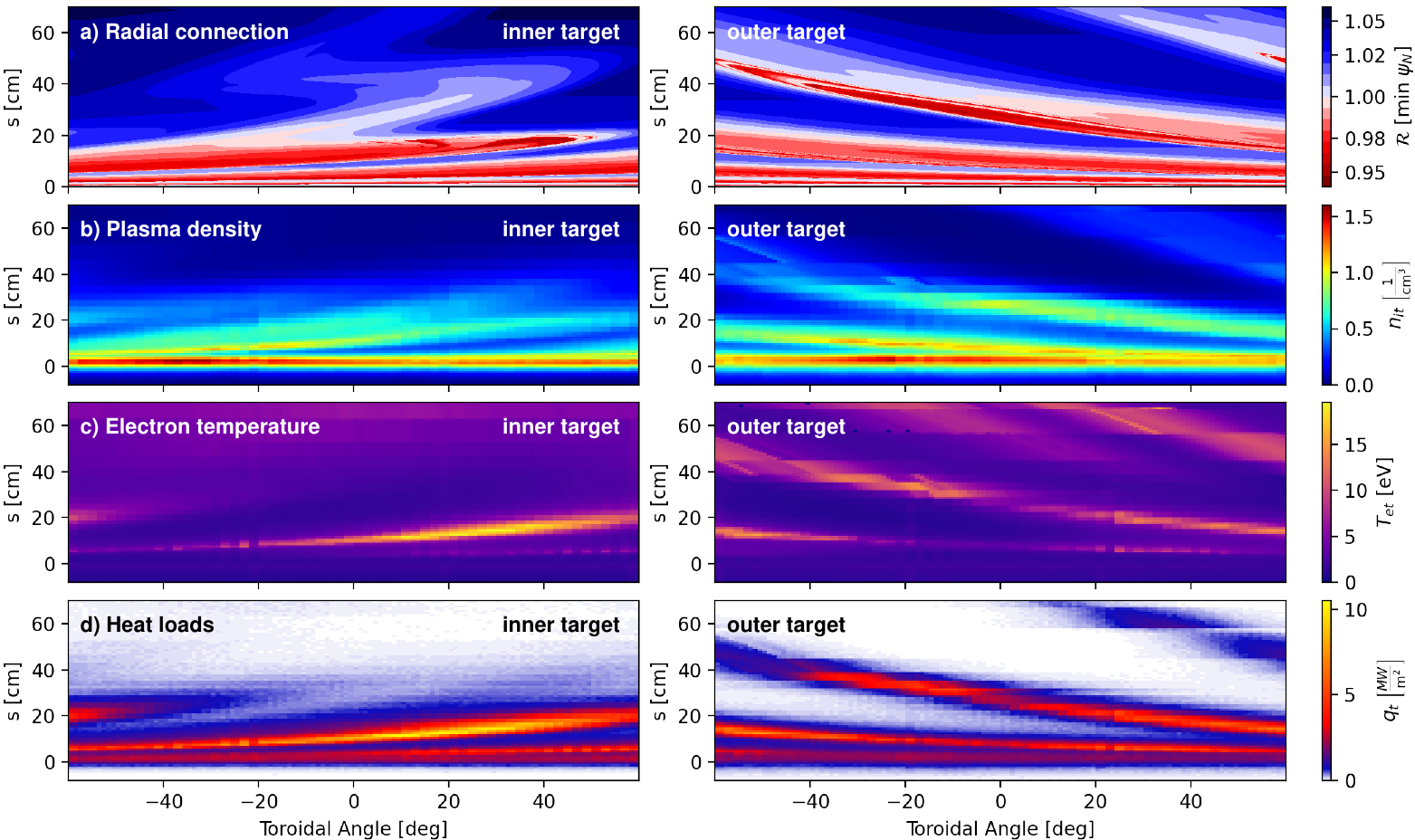}
\caption{Magnetic footprint and plasma conditions on the inner divertor target (left column) and outer divertor target (right column) for the \CmaxX configuration at $\IRMP \, = \, 60 \, \kAt$: (a) radial field line connection, (b) plasma density, (c) electron temperature, and (d) heat loads.}
\label{fig:footprint_xiXmax60}
\end{figure*}

A larger magnetic footprint is expected for the higher $\IRMP \, = \, 60 \, \kAt$ of the conservative scenario.
Figure \ref{fig:footprint_xiXmax60} shows the resulting impact on plasma conditions at the divertor targets.
Conditions near the equilibrium strike point are comparable to figure \ref{fig:footprint_xiXmax45}, but the impact of the lobes reaches further out into the SOL.
Higher temperatures of $15 - 20 \, \electronvolt$ can be found with peak heat loads of around $10 \, \mega\watt \, \meter^{-2}$ on the inner divertor target - just around the limit of acceptable values.
Those peaks result from heat flux that is routed into the helical lobes upstream, and very little dissipation from Ne impurities is found along the way to the target.
A combination of higher density from additional $D_2$ gas puffing and higher impurity concentration may mitigate the impact on the inner divertor target.
This will be done in future work.

\begin{figure}
\centering
\includegraphics[width=80mm]{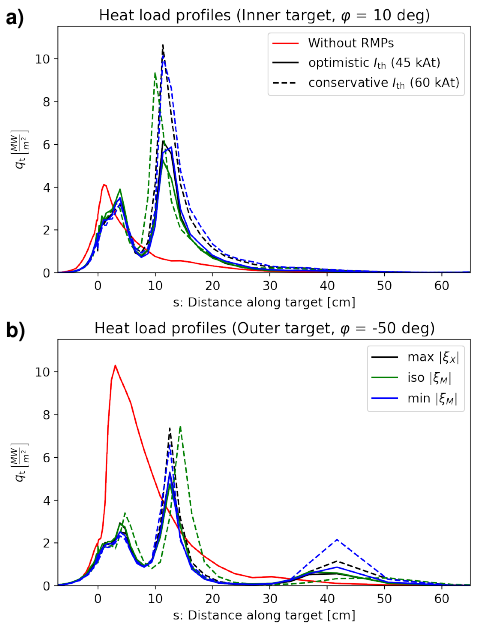}
\caption{Heat load profiles along the inner target at $\varphi = 10 \, \deg$ (a) and along the outer target at $\varphi = -50 \, \deg$ (b) for the \CmaxX (black), \CisoM (green) and \CminM (blue) RMP configurations for the optimistic ELM suppression threshold (solid) and the conservative one (dashed) in comparison to a simulation without RMPs (red).}
\label{fig:target_profiles}
\end{figure}

\begin{figure}
\centering
\includegraphics[width=80mm]{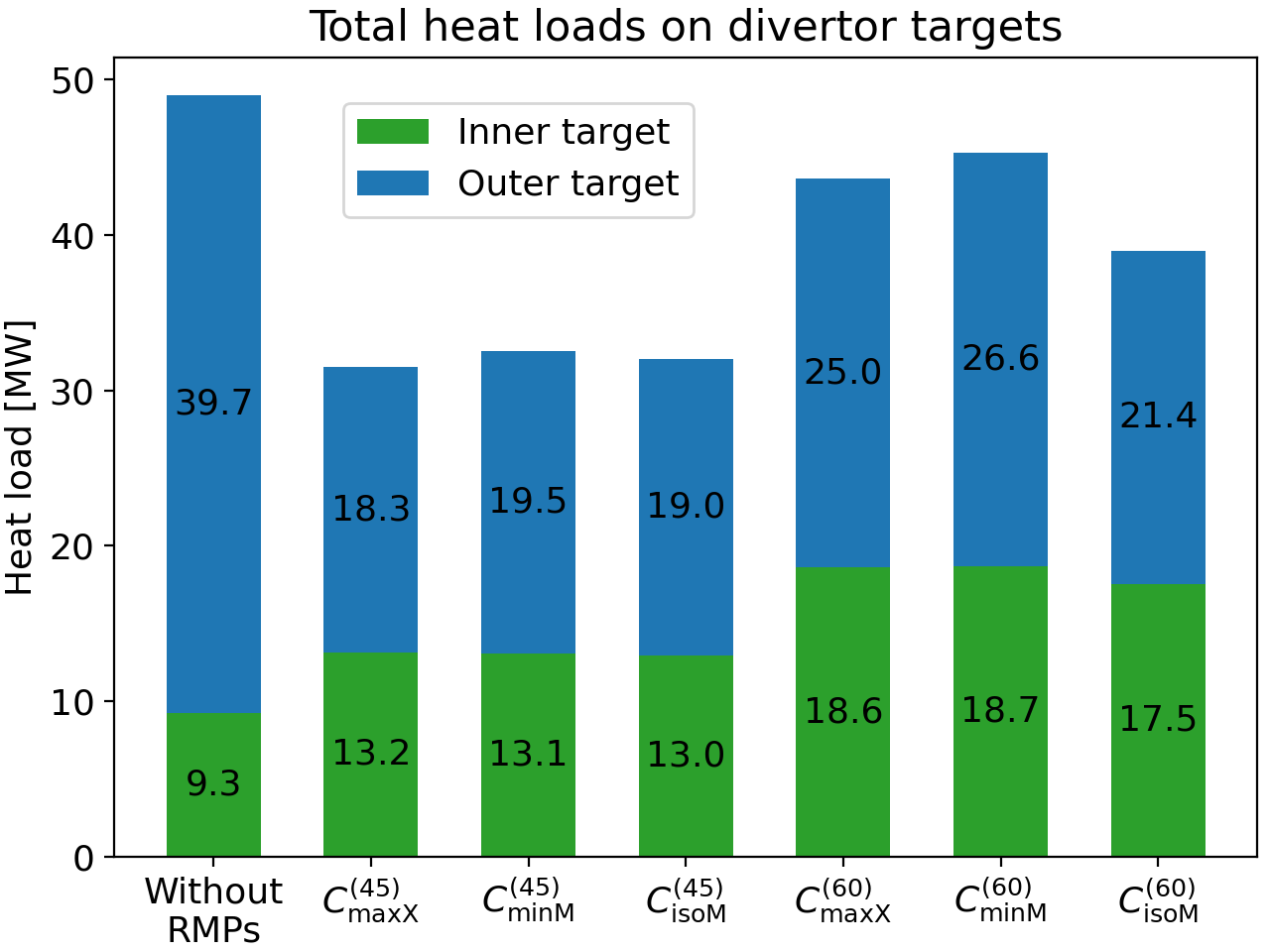}
\caption{Total head loads on inner and outer divertor targets for all simulations.}
\label{fig:power_balance}
\end{figure}

Figure \ref{fig:target_profiles} shows a comparison of heat load profiles for the 3 different RMP configurations for both the optimistic ELM suppression threshold (solid) and the conservative one (dashed) in comparison to a simulation without RMPs (red). 
Without RMPs, the plasma at the inner target is detached with a peak heat load of $4 \, \mega\watt \, \meter^{-2}$ but remains attached at the outer target at $10 \, \mega\watt \, \meter^{-2}$ for the model parameters discussed above.
The total head loads are summarized in figure \ref{fig:power_balance}, which shows that only $9.3 \, \mega\watt$ are deposited on the inner target while $39.7 \, \mega\watt$ are deposited on the outer target without RMPs.
With RMPs, dissipation from impurity radiation is less effective in inner divertor, which leads to higher peak and integral values in figure \ref{fig:target_profiles} (a) and \ref{fig:power_balance}, respectively.
In the outer divertor, on the other hand, impurity radiation becomes more effective with RMPs, which leads to a lower integral heat load.
The main heat load peak near the equilibrium strike point is reduced to about $2 \, \mega\watt \, \meter^{-2}$, but at the expense of secondary peaks of $5 -- 7 \, \mega\watt\, \meter^{-2}$ further away at $s = 12 -- 15 \, \centi\meter$.
Tuning of the RMP spectrum appears to have marginal impact on the heat loads for the same \xiXnorm, although a small reduction of the peak heat load on the inner target can be found for the \CisoM configuration (green).
In any case, the simulations show that both optimistic and conservative ELM suppression scenarios are (marginally) compatible with divertor limits - at least for the selected set of model parameters.

\section{Conclusions}

A conservative and an optimistic subspace of RMP configurations for ELM suppression has been predicted for H-mode burning plasmas at 15 MA current and 5.3 T magnetic field in ITER. 
This is based on the X-point displacement \xiXnorm from linear MHD plasma response (MARS-F) as figure of merit for edge stability, which is then compared to a threshold value.
Based on published results from non-linear MHD modeling (JOREK), this threshold value is taken from a particular RMP configuration that shows onset of ELM suppression in the range of $45 \, \kAt$ (optimistic) to $60 \, \kAt$ (conservative).
The plasma boundary displacement at the outboard midplane has been used as a proxy for perturbation of the core, and it has been shown that it can be reduced by a factor of 2 for equivalent edge stability proxy.
At the same time, variations of the magnetic footprint size within the ELM suppression subspace remain within $20 \, \%$.
It has been demonstrated that the predictions of the RMP subspace for ELM suppression are robust by comparing to figures of merit based on the kink amplitude and on the resonant response.
Those are also consistent with the RMP subspace for ELM suppression in KSTAR.

The magnetic footprint of perturbed field lines connecting from the main plasma (normalized poloidal flux $< 1$) to the divertor targets is found to be significantly larger than the expected heat load width in the absence of RMPs.
This facilitates heat load spreading with peak values at an acceptable level below $10 \, \mega\watt \, \meter^{-2}$ on the outer target already at moderate gas fueling and low Ne seeding for additional radiative dissipation of the $100 \, \mega\watt$ of power into the scrape-off layer (SOL).
On the inner target, however, re-attachment is predicted away from the equilibrium strike point due to the heat flux that is routed into the helical lobes upstream, but with less efficient impurity radiation at higher downstream temperatures.
Tuning of the RMP spectrum appears to have marginal impact on the heat loads for the same \xiXnorm.
In any case, the simulations show that both optimistic and conservative ELM suppression scenarios are (marginally) compatible with divertor limits - at least for the selected set of model parameters.
Future work is planned to explore the impact of higher density from additional $D_2$ gas puffing and higher impurity concentration on the heat load peaks away from the equilibrium strike point.


\appendix
\section*{Acknowledgments}
This work was supported by the U.S. Department of Energy under award No. DE-SC0020284 and DE-SC0020357.
This work was done under the auspices of the ITER Science Fellow Network. The views and opinions expressed herein
do not necessarily reflect those of the ITER Organization.

\section*{References}
\bibliographystyle{unsrt_doilink}
\bibliography{references}

\end{document}